\documentstyle[ltwol,epsf]{article}

\def\be{\begin{equation}}
\def\ee{\end{equation}}
\def\bea{\begin{eqnarray}}
\def\eea{\end{eqnarray}}

\def\xba{\bar}

\begin{document}

\title{
Strategies for The Determination of $\phi_3$ in $B^-\to D^0 K^-$
}

\author{D. Atwood}

\address{
Dept. of Physics, Iowa State University, Ames, IA 50014, USA\\
E-mail: atwood@iastate.edu
}

\author{\it Submitted to the proceedings of the Fourth International
Conference on CP Violation in B Physics}


\twocolumn[\maketitle\abstracts{ 
Direct CP violation in decays such as $B^- \to D^0 K^-$ is sensitive to
the CKM angle $\phi_3$ because these decays allow the interference of
$b$-quark to $c$-quark with $b$-quark to $u$-quark transitions.  Indeed,
$\phi_3$ may be determined if one can infer the strong phase of the $B$
and subsequent $D^0$ decays from experimental data. In this talk, I will
discuss how this can be carried out using either a single decay mode of
the $D^0$ by combining data from a number of $D^0$ decay modes as well as
the use of other, analogous decays and the prospects of implementing such
methods at various $B$-factories.
Since the properties of the $D^0$ decays are crucial to these methods, it
is possible that $D^0$-$\overline {D^0}$ mixing at the 1\% level will
contaminate the results. I will therefore discuss various methods to
remove such confounding effects so that $\phi_3$ may be determined even if
such mixing is present.
}]

\section{Introduction}

The asymmetric B experiments BaBaR~\cite{babar_ref} and
BELLE~\cite{belle_ref} have already obtained preliminary measurements of
the angle $\phi_1$ of the unitarity triangle~\cite{utriang} of the Cabibbo
Kobayashi-Maskawa~\cite{ckmref} (CKM) matrix through the ``gold-plated''
mode~\cite{psiks} $\psi K$. Using $B^0\xba B^0$ oscillation it may also be
possible to extract $\phi_2$ via modes such as $\pi\pi$ and
$\pi\rho$~\cite{alpharef} however to extract $\phi_3$ using oscillations
requires $B_s$-mesons and so is inaccessible to $\Upsilon(4S)$ machines. 

Although experiments to extract $\phi_3$ via $B_s$ oscillations may be
performed at hadronic $B$-facilities it is also possible to measure
$\phi_3$ through direct CP violation in the $B$ system. Thus, the complete
set of unitarity angles may, in principle, be accessible at $\Upsilon(4S)$
machines. Specifying as many parameters of the unitarity triangle as
possible is, of course an important check of the Standard Model (SM). In
addition, if $\phi_3$ is measured via direct CP violation 
the comparison to the measurement through indirect
CP-violation in the $B_s$ system provides another non-trivial check of the
SM.

The idea behind the measurement of $\phi_3$ through direct CP violation is
to consider a process which allows interference of the quark level
processes $b\to \xba u c s$ and $b\to u \xba c s$.  This may be
accomplished if both processes ultimately hadronize to a common final
state~\cite{isi,glw,ads1,ads2}. In particular $b\to \xba u c s$ can drive
the decay $B^-\to D^0 K^-$ while $b\to u \xba c s$ can drive the decay
$B^-\to \xba D^0 K^-$ which will thus interfere provided that $D^0$ and
$\xba D^0$ are detected through decays into a common final state.

In this talk, I will discuss various strategies for the determination of
$\phi_3$ in such decays.  The most crucial element is the selection of the
$D^0$ decay modes which are to be used.  In the simplest case is where CP
violation is seen in a single mode we will see that there is not enough
information to precisely determine $\phi_3$. However, in Section~2 I will
show if the CP violation is large, significant bounds may be placed on
$\phi_3$ thus modes where CP violation may be large are especially
significant.  More generally, as discussed in Section~3, $\phi_3$ may be
extracted if two or more modes are measured. A single three body mode may
also give equivalent information because each point on the $D$-decay Dalitz
plot may be regarded as a separate ``mode''. These methods are subject to
possible contamination from $D\bar D$ oscillation if it is on the order of
$1\%$, I will discuss the impact of this possibility and methods to deal
with it in Section~4 and in Section~5 I will give my conclusions.

\section{One $D^0$ Decay Mode}

Consider the case where $D^0$ and $\xba D^0$ can decay to a common final
state $X$. This may either be CP eigenstate (e.g. $K^+K^-$) or a state
such as $K^+\pi^-$ which is a Cabibbo allowed (CA) decay of the $\xba D^0$
but a doubly Cabibbo suppressed (DCS) decay of the $D^0$. In either case,
one can determine $\phi_3$ if one can measure~\cite{glw} the rates
$d(X)=Br(B^-\to
K^- [X])$ and $\xba d(\xba X)=Br(B^+\to K^+ [\xba X])$ (here $[X]$ means a
decay to $X$ via $D^0$ mixed with the $\xba D^0$ channel) provided one
also knows the branching ratios $a=Br(B^-\to K^- D^0)$; $b=Br(B^-\to K^-
\xba D^0)$; $c(X)=Br(D^0\to X)$ and $\xba c( X)=Br(\xba D^0\to X)$.  This
information allows us to solve (up to an eight fold ambiguity) for the
weak phase $\phi_3$ as well as the total strong phase difference $\xi$.

In practice, however, $b$ is not easy to determined directly because it is
difficult to find a prominent tag for $\xba D^0$.  For instance, a
leptonic tag has a large background from leptonic decay of $B^-$ while if
one tries to tag it through decays such as $\xba D^0\to K^+\pi^-$ the
signal is subject to interference effects from $D^0\to K^+\pi^-$. In fact
it is the existence of such interference effects we wish to exploit as a
source of CP violation that make the direct determination of $b$ via a
hadronic tag impossible. 

Of course if CP violation were seen (i.e. $d\neq\xba d$) then within the
SM it must be the case that $\phi_3\neq 0$. As might therefore be
expected, in the absence of $b$ we can use the rest of the information to
establish a lower bound on $\phi_3$. In particular, if we define
$Q\equiv\sin^2\phi_3$ then we obtain the 
following bound on $Q$ and incidentally $b$:

\begin{eqnarray}
Q\geq Q_{min}= (1+z)\left(1-\sqrt{1-y/(1+z)} \right)/2\nonumber\\ 
-\sqrt{1+z+|y|} 
\leq 
\sqrt{u}-1
\leq
\sqrt{1+z-|y|}
\label{inequal}
\end{eqnarray}

\noindent
where
$1+z=(d(X)+\xba d(\xba X))/(2 a c(X))$, 
$y=(d(X)-\xba d(\xba X))/(2 a c(X))$ 
and
$u={b\xba c(X)/a c(X)}$.

In the case where $z$ is negative, an upper bound~\cite{gronau_private} 
can also be placed on 
$Q$:
\begin{eqnarray}
Q\leq 1+z
\end{eqnarray}

\noindent
which is similar in form to the upper bound on $Q$ obtained from 
tree-penguin interference in $B\to K\pi$~\cite{fl_man}. Note that it
can apply even if CP violation is absent however 
$z$ 
may only be negative if it happens that 
$u\leq 4$.

\begin{figure}
\epsfysize 2.5 in     
\mbox{\epsfbox{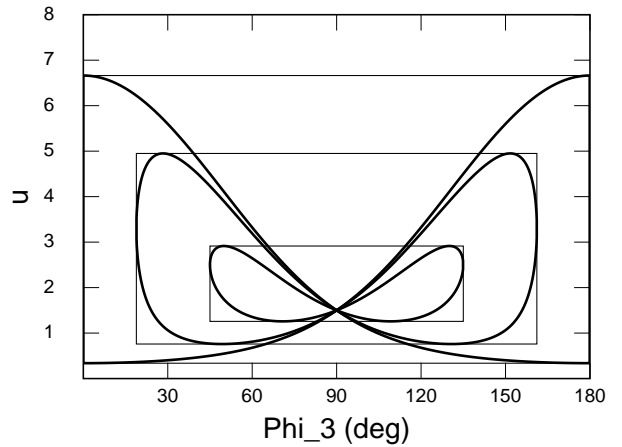}}
\caption{
Each of the solid lines shows the locus of points in $\phi_3$ versus $u$
of allowed solutions given $z_i=1.5$ for $y_i=0$ (outer curve), $1$
(intermediate curve) and $2$ (inner curve). The boxes indicate the
inequalities Eqn.~(\ref{inequal}). Note that $u\propto b$.
\label{figure1}
}
\end{figure}

Motivation for these bounds can be found in Fig.~\ref{figure1} where we
plot the
relation between $u$ and $\phi_3$ for the experimental inputs $z=1.5$ and
$y=0$, $1$ and $2$ where the larger values of $y$ correspond to the
smaller 
``lazy eight''
curves. The boxes indicate the bounds established by the
inequality eq.~(\ref{inequal}). Since $y$ is proportional to CP violation,
it is clear that the most stringent inequality bounds obtain where CP
violation is large. A useful property of these curves is that even though 
the strong phase difference is not explicitly given, it may be read off
(up to a four fold ambiguity) since for a given value of $(\phi_3,u)$ on
the curve, the horizontal line through that point also intersects the
curve at $(\xi^\prime,u)$ where one of 
$\{\xi^\prime$ ,$\pi-\xi^\prime$, $\pi+\xi^\prime$, $\xi^\prime\}$ is the
strong phase difference.

Large CP violation is only possible when the two interfering amplitudes
are similar in magnitude. Such a situation may happen if we consider a
final state $X$ 
where
$D^0\to X$ is DCS (e.g. $K^+\pi^-$). Thus, while
$a$ is about two orders of magnitude greater than $b$
due to color suppression, this if offset by
$\xba c(X)$ being about two orders of magnitude greater than $c(x)$ so
$b\xba c(X)\sim a c(X)$. It is obviously advantageous to experimentally
study all modes of this kind in order to find the one which gives the
largest lower bound on $Q$.

Of course we would like to determine $\phi_3$ exactly rather than just
establishing a bound for it. There are three possible approaches to
obtaining this quantity if $b$ 
cannot be experimentally measured: First of
all, one could use a theoretical model to estimate 
$b$.
Second, one could use an analogous decay $B^0\to K^0[\bar D\to K^+\pi^-]$
or
$\Lambda_b\to \Lambda [\bar D^0\to K^+\pi^-]$ where the cross channel is
is not color enhanced and interference are $\sim 30\%$ (we will discuss
this more below) or third of all, one can consider multiple $D$ decay
modes 
each of which may decay with a different strong phase.
In this last case I also include $D$ decays to a multi-body final
state where each point in phase space may be considered as a separate
``mode''.

\section{Two Modes and Three Body Decays of $D^0$}

If $d$ and $\xba d$ are measured for exactly two modes, then there are
the same number of equations as unknowns and $Q$ may be determined up to
some discrete ambiguity. Graphically, two general curves such as in
Fig.~\ref{figure1}
will intersect at up to 4 points so in this case there may be a 4-fold
ambiguity in $Q$ therefore a 16-fold ambiguity in $\phi_3$. If three or
more modes are considered then the curves in the $b-\phi_3$ plane should
only intersect at one point in the first quadrant of $\phi_3$ so $\phi_3$
has a 4-fold ambiguity.

\begin{figure}
\epsfysize 3.0 in     
\epsfbox{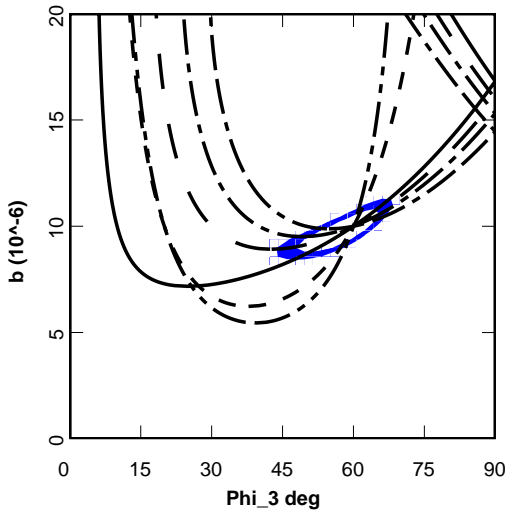}
\caption{
The curves in the $\phi_3-b$ plane 
using the parameters considered 
in~$^{10}$
are shown for the modes 
$K^+\pi^-$ (solid curve);   
$K_S\pi^0$ (short dashed curve);
$K^+\rho^-$ (long dashed curve);
$K^+a_1^-$ (dash-dot curve);
$K_s \rho^0$ (dash-dot-dot curve)
and
$K^{*+} \pi^-$ (dash-dash-dot curve). 
The inner edge of the shaded region
corresponds to the 
$68\%$ 
CL for $\hat N=10^8$ while the outer edge
corresponds to the 
$95\%$ 
CL.
\label{figure2}
}
\end{figure}

In Fig.~\ref{figure2} we show the results of a sample calculation
from~\cite{ads2}
where the modes:  $K^+\pi^-$, $K_S\pi^0$, $K^+\rho^-$, $K^+a_1^-$,
$K_S\rho^-$, $K^{*+}\pi^-$ were considered. For the parameters of that
calculation, the inner edge of the shaded region indicates the $68\%$ CL
and the outer region indicates the $95\%$ CL given $\hat N=({
number~of~B^\pm})({acceptance})$ = $10^8$. In this example it was found
that
with $\hat N=10^8$ statistical errors in $\phi_3$ were $\sim
5^\circ-10^\circ$ for
a variety of initial values of $\phi_3$ and strong phases.

In order to determine $\phi_3$ is is therefore advantageous to consider a
number of modes. In addition to the different final states 
such as those
considered
above, we can also replace $K\to K^*$ and $D\to D^*$. Of course if we do
both at once so that both sides have $J\neq 0$ then we need to do a more
complicated angular analysis as considered by~\cite{sinhas}.  Note that
here
$\phi_3$ is common but each case has a separate $b-axis$. To Drive up
additional modes we can also consider analogous decays where we replace
the spectator with a $\bar d$ 
(i.e. $B^0\to D^0K^0$)
or a $ud$ (i.e. $\Lambda_b\to D^0\Lambda$). 
The point here is that we may be
justified in putting these cases on a common $b$ axis. 

Because the main point of combining multiple modes is to overcome the lack
of knowledge of $b$, $B^0\to K^0 D$ where the $D$ is decays to a
state such as $K^+\pi^-$ may be a particularly important mode to use in
this way since the dominant contribution is proportional to $b$ while $a$
is color suppressed in this channel~\cite{wip}. 
As emphisized by~\cite{adsXglw}, the complimentary case where 
$B^-\to K^- D^0$ and
$D^0$ decays
to a CP eigenstate (e.g. $\pi^+\pi^-$) and so the $a$ channel is much
larger than the $b$ channel
also gives the same kind of
information since 
the amount of interference evident in the system determines $b$ without
a strong dependence on $\phi_3$.


An additional source of constraints that can be helpful may be obtained
from a charm factory data which can constrain the strong phase differences
between $D^0$ and $\xba D^0$ decays as well as give definitive information
concerning $D\xba D$ oscillations~\cite{sofer_charm,gronau_charm}.

Note that a number of the modes we consider are instances of three body
final states. For a single three body mode (e.g. $K^+\pi^-\pi^0$) we can
consider each of the points on the dalitz plot as a having a separate
strong phase so clearly in principle there is enough information to
determine $\phi_3$. One can thus fit the data to a resonance model as
in~\cite{e687} together with the overall strong and weak phase
differences.

In this talk I would like to emphasize a different method of analysis
based on the saturation of eq.~(\ref{inequal}). 
Regarding each point of the Dalitz plot as a separate mode, one may find
the value of $Q_{min}$ in Eqn.~(\ref{inequal}) for each point of the
Dalitz plot. Just as in the case of a number of discrete modes, the true
value of $Q$ must exceed all lower bounds.  In this case, however, because
the strong phases due to resonances vary across the Dalitz plot, it is
likely that the greatest value of $Q_{min}$ is in fact equal to $Q$. 

\begin{figure}
\epsfysize 2.5 in     
\epsfbox{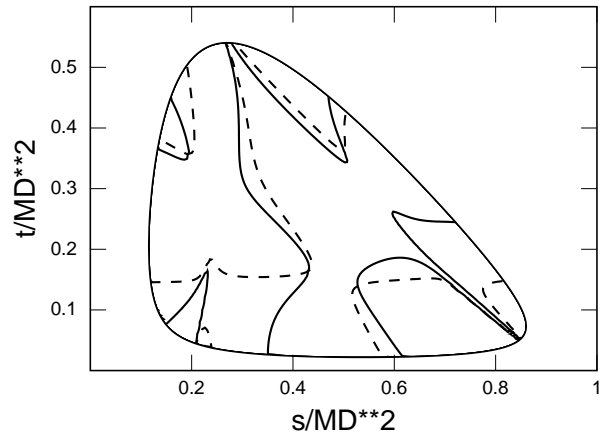}
\caption{
The locus of points on a dalitz plot 
for the final state $K^+\pi^-\pi^0$
where $Q_{min}=Q$ for
$\phi_3=60^\circ$ and 
an overall strong phase difference of 
$0^\circ$ (solid line)
and 
$\zeta=60^\circ$ (dashed line). 
Here 
$s=(p_{\pi-}+p_{K+})^2$
and
$t=(p_{\pi-}+p_{\pi 0})^2$.
\label{figure3}
}
\end{figure}

In Fig.~\ref{figure3} I show a map ``magic'' points where $Q=Q_{min}$ on
the Dalitz plot for the case of $D^0\to K^+\pi^-\pi^0$ using the model
of~\cite{ads2} that uses the data from E687~\cite{e687} as input together
with SU(3) to give the DCS channel. Here I have taken $\phi_3=60^\circ$
with an overall strong phase difference of $0^\circ$ for the solid curve
and $60^\circ$ for the dashed curve.

\section{The Influence of $D\xba D$ mixing}

In the discussion so far we have explicitly assumed that $D^0\xba D^0$ was
negligible. In particular, since we often take advantage of interferences
involving DCS decays which are $O(1\%)$ of the interfering CA decay, the
total probability of mixing must be less than $O(1\%)$ for this case to
remain valid.  In fact, it has been suggested that the standard model may
cause $D\xba D$ mixing at about this level~\cite{ddmix_resonance}. In
fact, it can be shown~\cite{ads2} that the changes to $d$ and $\xba d$
from such mixing will be $O(10\%)$ which leads to an inherent error in the
determineation of $\phi_3$ of $\sim
10^\circ-15^\circ$~(\cite{soffer_silva}). 

In order to overcome this possible systematic error, there are two
approaches:

\begin{enumerate}

\item Using information on the the time between the $B^-$ decay and the
subsequent $D^0$ decay, then the effects of possible mixing can be
eliminated. 

\item
If the parameters of $D\xba D$ mixing are known independently, then they 
can be taken into account in interpreting the time integrated data

\end{enumerate}

\noindent
Indeed, if the mixing parameters and time dependent data is available, 
then one can in principle extract $\phi_3$ from just one mode though most 
likely, the time dependence in the decay is too weak to make this a 
useful method. 

Here I will emphasise the fact that if we have time dependent data, 
it is particularly simple to separate out the contributions of mixing and
thus proceed with the analysis as in the absence of mixing at some cost in
statistics.

The key point is that for $D^0$ the decay time is much shorter than the
oscillation time and therefore it is valid to write

\begin{eqnarray}
{d\over d\tau}d(X)&\approx&(d_0(X)+d_1(X)\tau)e^{-\tau}
\nonumber\\
{d\over d\tau}\xba d(\xba X)
&\approx& 
(\xba d_0(\xba X)
+
\xba d_1(\xba X)\tau)e^{-\tau}
\end{eqnarray}

\noindent
where $\tau=t\Gamma_D$. Thus $d_0(X)$ and $\xba d_0(\xba X)$ would be the
branching ratios absent mixing so if we extract them from the time
dependence we may proceed as if there were no mixing.

This can be accomplished through weighting the data with $w_0(\tau)  =
2-\tau$
so that

\begin{eqnarray}
d_0=
\int_0^\infty	 
[d(\tau)w_0(\tau)] d\tau 
;
~~~
\xba d_0=
\int_0^\infty
[\xba d(\tau)w_0(\tau)] d\tau
\label{timeweight}
\end{eqnarray}

Using this method more data would be required to obtain the same
statistical results as in the unmixed case. In the unmixed case where
$d_1=0$ one could obtain $d_0$ more effectively by taking the time
integrated rate. Thus in the unmixed case if a measurement of $d_0$ is
based on $n$ events, the uncertainty in $d_0$ is given by: 
${(\Delta d_0)^2/ d_0^2}
=
{1/ n}$.
In the mixed case, using eqn.~(\ref{timeweight}) the uncertainty
is
${(\Delta d_0)^2/ d_0^2}
=
(2/ n)\left(1+{d_1/ d_0}\right)^2$.
Since $d_1\sim O(d_0/10)$, this
means that roughly twice the
data is needed to have the same statistical power as in the unmixed case. 
In order to gauge the precision  
of time measurement required, we can smear out the distribution in 
eq.~(\ref{timeweight}) with a Gaussian resolution function of the form 
$r(\tau,\tau^\prime)\propto e^{-{(\tau-\tau^\prime)^2\over 2\sigma^2}}$
where $\tau$ is the actual time of the decay, $\tau^\prime$ is
the measured time of the decay and $\sigma$ is the resolution (all in
units of $1/\Gamma_D$).  Since $r$ is symmetric under $\tau\leftrightarrow
\tau^\prime$, the fact that $w$ is linear in $\tau$ implies
eq.~(\ref{timeweight}) will still be true for $\tau^\prime$ but now the
error is: 

\begin{eqnarray}
{(\Delta d_0)^2/ d_0^2}
=
({2/ n})(1+\sigma^2)\left(1+{d_1/ d_0}\right)^2 
\end{eqnarray}

\noindent
As can be seen, the number of events required is not adversely effected if 
$\sigma \leq 1/\Gamma_D$ but will be significantly degraded otherwise.

\section{Conclusions} 

In conclusion, in the $B^\pm$ system direct CP violation in the decay $D^0
K^\pm$ may provide a means of determining $\phi_3$ with $10^8-10^9$ $B$
mesons. The key is to observe the correct $D^0$ or combination of $D^0$
decay modes. If large CP violation is seen in any one mode, it may
establish a lower bound on $\sin^2\phi_3$ while data from multiple modes
or three body modes can be used to determine $\sin^2\phi_3$.

\section*{Acknowledgments}

I would like to acknowledge useful discussions with M.~Gronau, D.~London
and R.~Sinha.
This research was supported in part by US DOE Contract Nos.\
DE-FG02-94ER40817 (ISU).


\section*{References}
\begin{thebibliography}{99}



\bibitem{babar_ref}
B.~Aubert {\it et al.}  [BABAR Collaboration],
Phys.\ Rev.\ Lett.\ {\bf 86}, 2515 (2001).



\bibitem{belle_ref}
A.~Abashian {\it et al.}  [BELLE Collaboration],
Phys.\ Rev.\ Lett.\ {\bf 86}, 2509 (2001).



\bibitem{utriang}
L.~L.~Chau and W.-Y. Keung, Phys.\ Rev.\ Lett.\ {\bf
53}, 1802 (1984).


\bibitem{ckmref}
N.~Cabibbo, Phys.\ Rev.\ Lett.\ {\bf 10}, 531 (1963);
M.~Kobayashi and T.~Maskawa, Prog.\ Th.\ Phys.\ {\bf
49}, 652 (1973).  




\bibitem{psiks}
I.I.~Bigi and A.I.~Sanda,
Nucl. Phys. {\bf B193}, 85 (1981);
Nucl. Phys. {\bf B281}, 41 (1987). 





\bibitem{alpharef}
M.~Gronau and D.~London, Phys. Rev.
Lett. {\bf 65}, 3381 (1990);
A.~E.~Snyder and H.~R.~Quinn,
Phys.\ Rev.\ D {\bf 48}, 2139 (1993).

\bibitem{isi}
I.~Dunietz, Phys. Lett. {\bf B270}, 75 (1991);
I.~Dunietz, Z.~Phys. {\bf C56}, 129 (1992);
I.~Dunietz, article in 
{B Decays}, S.~Stone ed. (World Scientific, Singapore,
1992). 




\bibitem{glw}
M.~Gronau and D.~Wyler, Phys.\ Lett.\ {\bf B265}
177 (1991); M.~Gronau and D.~London., Phys.\ Lett.\ {\bf B253}, 483
(1991).




\bibitem{ads1}
D.~Atwood, I.~Dunietz and A.~Soni, Phys.\ Rev.\ Lett.\ {\bf
78}, 3257 (1997). 


\bibitem{ads2}
D.~Atwood, I.~Dunietz and A.~Soni,
Phys.\ Rev.\ D {\bf 63}, 036005 (2001).

\bibitem{gronau_private}
M.~Gronau,
hep-ph/0001317; M.~Gronau, private communication.

\bibitem{fl_man}
R.~Fleischer and T.~Mannel,
Phys.\ Rev.\ D {\bf 57}, 2752 (1998).


\bibitem{sinhas}
N.~Sinha and R.~Sinha, Phys.\ Rev.\ Lett.\ {\bf 80}, 3706
(1998).

\bibitem{wip}
D.~Atwood, {\it work in progress}.


\bibitem{adsXglw} 
A.~Soffer, Phys.\ Rev.\ {\bf D60}, 054032 (1999). 




\bibitem{sofer_charm}
A.~Soffer, hep-ex/9801018.


\bibitem{gronau_charm}
M.~Gronau, Y.~Grossman and J.~L.~Rosner,
hep-ph/0103110.


\bibitem{e687} P.~L.~Frabetti {\it et al.} [E687 Collab.], Phys.\ Lett.\
{\bf B331},
217 (1994). 


\bibitem{ddmix_resonance}
M.~Gronau,
Phys.\ Rev.\ Lett.\ {\bf 83}, 4005 (1999).


\bibitem{soffer_silva}
J.~P.~Silva and A.~Soffer, Phys.\ Rev.\ {\bf D61},
112001 (2000). 






\end{thebibliography}
\end{document}